\documentclass[aps,prb,superscriptaddress,amsmath,amssymb,floatfix,twocolumn]{revtex4}
\usepackage{graphicx}
\usepackage{subfigure}
\usepackage{color}

\begin{document}

\title{ Spin Dynamics of a $J_1-J_2$ antiferromagnet and its implications for iron pnictides}

\author{Pallab Goswami} \affiliation{Department of Physics and Astronomy, Rice University, Houston,
TX 77005}
\author{Rong Yu} \affiliation{Department of Physics and Astronomy, Rice University, Houston,
TX 77005}
\author{Qimiao Si} \affiliation{Department of Physics and Astronomy, Rice University, Houston,
TX 77005}
\author{Elihu Abrahams} \affiliation{Department of Physics and Astronomy, University
 of California Los Angeles, Los Angeles, California 90095}

\begin{abstract}
Motivated by the recent observation of antiferromagnetic correlations
in the paramagnetic phase of iron pnictides, we study the finite-temperature spin dynamics of a two-dimensional $J_1-J_2$ antiferromagnet.
We consider the paramagnetic phase in the
regime of a $(\pi,0)$ collinear ground state,
using the modified spin wave theory. Below the mean-field Ising
transition temperature,
we identify short-range anisotropic antiferromagnetic correlations.
We show that the dynamical
structure factor $\mathcal{S}(\mathbf{q},\omega)$
contains elliptic features in the momentum space,
and determine its variation with temperature and energy.
Implications for the spin-dynamical experiments in the iron pnictides are discussed.
\end{abstract}

\maketitle

\section{Introduction}
High-temperature superconductivity in the iron
pnictides \cite{Kamihara_FeAs,Zhao_Sm1111_CPL08}
arises by doping antiferromagnetic  parent compounds \cite{Cruz}.
Hence, the strength of the electronic correlations,
the nature of magnetism, and the relationship between magnetic
excitations
and the superconductivity
are
important issues for understanding the emergence of high temperature
superconductivity in these materials.
In the parent iron pnictides,
the N\'eel transition into a $(\pi,0)$ antiferromagnet
is either preceded by or concomitant with
a tetragonal-to-orthorhombic
structural transition.
The $(\pi,0)$ magnetic order by itself can be understood either
by invoking a local moment $J_1-J_2$
model \cite{Si, Yildirim,Ma, Fang:08, Xu:08, Si_NJP,Dai, Uhrig}
or an itinerant model with nearly nested electron
and hole pockets \cite{Graser, Ran, Knolle}.

The experimentally observed ``bad metal" behavior, the Drude-weight
suppression \cite{Qazilbash, Hu} and the temperature-induced spectral-weight
transfer \cite{Hu, Yang, Boris} place these materials near to
a Mott transition \cite{Si, Haule, Si_NJP,Kutepov10};
a Mott insulator can emerge when the iron square lattice either expands\cite{Zhu} or contains ordered vacancies \cite{MFang}.
In a metallic system close to a Mott transition, quasi-local moments are expected to arise;
this picture is further supported by the experimental observation
of zone boundary spin wave excitations in the magnetically
ordered state at low temperatures \cite{Zhao}.
The inelastic neutron scattering experiments demonstrated
the need for an anisotropic $J_1-J_2$ model with $J_{1x}\neq J_{1y}$,
which may reflect an orbital ordering \cite{Singh, Jaanen,Lv}
while pointing to
the relevance of magnetic frustration from the extracted ratio
$(J_{1x}+J_{1y})/2J_2 \sim 1$ \cite{Zhao}.
Therefore,
results in the tetragonal, paramagnetic phase of the parent compounds
are of great importance for understanding the relevance of an isotropic
$J_1-J_2$ model as well as the strength of the underlying
magnetic frustration. Recent inelastic neutron scattering measurements of
Diallo {\it et al.} \cite{Diallo} on the tetragonal, paramagnetic phase
of $\mathrm{CaFe_2As_2}$ represent a first step in this direction.
Even above the concomitant first-order structural and N{\' e}el transition
temperature, they have
observed anisotropic spin dynamics around the
$(\pi,0)$ wave vector, and
the inferred ratio
$J_1/J_2 \sim 0.55$ is similar to that of the ordered phase.

Motivated by these experimental results we study the spin
dynamics of a two-dimensional $J_1-J_2$ antiferromagnet.
While
theoretical studies exist
on the
order-from-disorder phenomenon and phase diagram
of the $J_1-J_2$
model \cite{Chandra,Flint},
the spin
dynamics in the paramagnetic phase
of the model
in the $(\pi, 0)$ collinear regime
has not yet been systematically studied.
We carry out the calculations
using a modified spin wave theory \cite{Takahashi1},
which incorporates the $1/S$ corrections that are important
for capturing the order-from-disorder phenomenon
and the associated dynamical properties. We
discuss the implications of our results for the iron pnictides,
including the role of
itinerant electrons.

Our paper is organized as follows. In Sec. II we introduce the relevant $J_1-J_2$ model and describe the modified spin wave theory calculations. In Sec. III we analyze the excitation spectrum obtained from modified spin wave theory, and associated behavior of the spin-spin correlation length. In Sec. IV we analyze the dynamic structure factor calculated by using the modified spin wave theory results. In Sec. V we consider the fluctuation effects due to itinerant electrons within a Ginzburg-Landau framework. In Sec. VI we describe the relation between our theoretical results and the experimental data obtained in the paramagnetic phase of iron pnictides. We provide a summary of our work in Sec. VII. The technical details of fitting the experimental data and consideration of inter-planar exchange coupling using modified spin wave theory are respectively relegated to Appendix A and Appendix B.

\section{Model and modified spin wave theory}
\begin{figure}[t!]
\centering\includegraphics[scale=0.3]{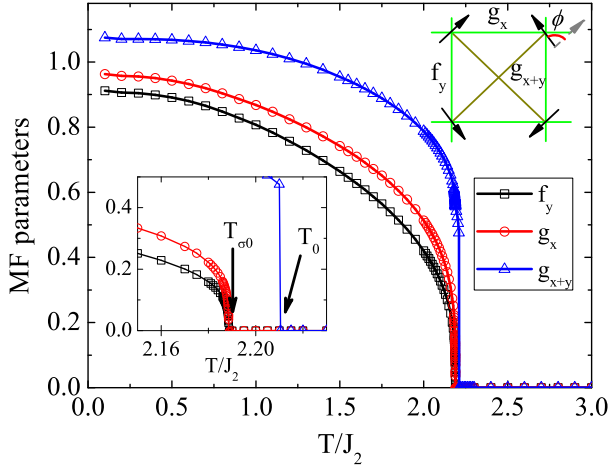}
\caption{(Color online)
The temperature dependence
of the mean-field parameters, for $S=1$ and $J_1/J_2=0.8$. The decoupled N{\'e}el sublattices are illustrated in the upper right corner, which also defines the angle $\phi$.
 }
\label{fig:1}
\end{figure}

The
model is defined by the Hamiltonian
\begin{equation}
H=J_1\sum_{\langle ij\rangle}\mathbf{S}_{i}\cdot \mathbf{S}_{j}
+J_2\sum_{\langle \langle ij\rangle \rangle}\mathbf{S}_{i}
\cdot \mathbf{S}_{j},
\end{equation}
where $J_1$ and $J_2$ respectively denote
the antiferromagnetic exchange couplings between spins
located in the nearest ($\langle ij\rangle$) and
next-nearest neighbor ($\langle \langle ij\rangle \rangle$)
sites on a square lattice.
Classically,
for $J_2/J_1>\alpha_c=0.5$,
the lattice
decouples into two independently N{\'e}el ordered, interpenetrating
lattices,
and the angle $\phi$ between
the staggered magnetizations of these two sublattices, as illustrated in the Fig.~\ref{fig:1} inset
is arbitrary.
An order-from-disorder transition
at temperature $T_{\sigma}$
breaks the fourfold rotational symmetry of the square lattice
down to a twofold rotational symmetry of the rectangular lattice,
and $\phi=0, \pi$ emerge as degenerate ground states at $T=0$
\cite{Chandra}.
Since quantum fluctuations make $\alpha_c>0.5$,
for definiteness we will focus on
$J_2/J_1 > 1$.

We define a local spin quantization
axis along the classical ordering direction at each site
($\boldsymbol \Omega_{i}^{cl}$),
as illustrated
in an inset to Fig.~\ref{fig:1}.
We then introduce the corresponding Dyson-Maleev (DM)
boson representation for the spin operators
at each site:
$\mathbf{S}_i\cdot \boldsymbol \Omega_{i}^{cl}=S-a_{i}^{\dagger}a_{i}$,
as well as $\mathbf{S}_{i}^{+}=\sqrt{2S}
(1-a_{i}^{\dagger}a_{i}/2S)a_i$
and $ \mathbf{S}_{i}^{-}=\sqrt{2S}a_{i}^{\dagger}$.
The modified spin wave theory \cite{Takahashi1}
treats the self-energy
of the $a$-bosons as a static quantity, which renormalizes
their dispersion;
in this respect, it is similar to
the large-$N$ Schwinger boson mean-field theory \cite{Arovas}.
Following Takahashi \cite{Takahashi1,Nishimori},
we express the Hamiltonian,
Eq.\ (1),  in terms of the DM bosons in momentum space.  The procedure is to minimize the free energy $\mathcal{F}=\langle H \rangle  -T\mathcal{S}$ under the constraint of zero magnetization, $\langle S - a_i^{\dag}a_i \rangle = 0$, with respect to variational parameters which enter ${\cal F}$.  These are the boson dispersion $\epsilon_{\bf k}$, the angle $\phi$ and the Bogoliubov angle $\theta_{\bf k}$. The latter enters in a Bogoliubov transformation that mixes the operators of the two interpenetrating N{\'e}el sublattices and renders the nonzero temperature density matrix diagonal \cite{Takahashi1}. The equal time correlators $\langle \mathbf{S}_{i}\cdot \mathbf{S}_{j}\rangle$ can be written in terms of expectation values like $\langle a_i^{\dag}a_j\rangle$. Therefore, we define ferromagnetic and antiferromagnetic bond correlations $f_{ij} = \langle a_{i}^{\dagger}a_j\rangle
=\langle a_{i}a_{j}^{\dagger}\rangle$ and
$g_{ij} = \langle a_{i}a_{j}\rangle=\langle a_{i}^{\dagger}a_{j}^{\dagger}\rangle$. The explicit expressions for the bond correlations are given by
\begin{eqnarray}
f_{ij}=\frac{1}{N}\sum_{\mathbf{k}}\cosh 2\theta_{\mathbf{k}}(n_{\mathbf{k}}+\frac{1}{2})\exp (-i\mathbf{k}\cdot \mathbf{r}_{ij}) \\
g_{ij}=\frac{1}{N}\sum_{\mathbf{k}}\sinh 2\theta_{\mathbf{k}}(n_{\mathbf{k}}+\frac{1}{2})\exp ( -i\mathbf{k}\cdot \mathbf{r}_{ij}),
\label{eq:FAFcorrelators}
\end{eqnarray}
where $n_{\mathbf{k}}=[\exp(\varepsilon_{\mathbf{k}}/T)-1]^{-1}$ is the Bose occupation factor.

In terms of $f_{ij}$ and $g_{ij}$ the equal time spin correlator $\langle \mathbf{S}_{i}\cdot \mathbf{S}_{j}\rangle$ can be expressed as
\begin{eqnarray}
\langle \mathbf{S}_{i}\cdot \mathbf{S}_{j}\rangle=\cos^2 \frac{\phi_{ij}}{2}\left[S+\frac{1}{2}-f(0)+f_{ij}\right]^2 \nonumber \\-\sin^2
\frac{\phi_{ij}}{2}\left[S+\frac{1}{2}-f(0)+g_{ij}\right]^2
\end{eqnarray}
where $\phi_{ij}=\phi, \ \pi- \phi, \ \pi$, for horizontal, vertical and diagonal bonds respectively (see Fig.~\ref{fig:1}). Using the expression for $\langle \mathbf{S}_{i}\cdot \mathbf{S}_{j}\rangle$ for different bonds, the total energy can be written as
\begin{widetext}
\begin{eqnarray}
&&E=\frac{J_1N}{2}\sum_{\boldsymbol \delta_1=\pm \hat{x}}\bigg[\cos^2 \frac{\phi}{2}\bigg(S+\frac{1}{2}-f(0)+f_x\bigg)^2- \sin^2
\frac{\phi}{2}\bigg(S+\frac{1}{2}-f(0)+g_x\bigg)^2\bigg]+\frac{J_1N}{2}\sum_{\boldsymbol \delta_2=\pm \hat{y}}\bigg[\sin^2
\frac{\phi}{2}\times \nonumber \\&&\bigg(S+\frac{1}{2}-f(0)+f_y\bigg)^2- \cos^2
\frac{\phi}{2}\bigg(S+\frac{1}{2}-f(0)+g_y\bigg)^2\bigg]-\frac{J_2N}{2}\sum_{\boldsymbol \delta_3=\pm \hat{x}\pm
\hat{y}}\left(S+\frac{1}{2}-f(0)+g_{x+y}\right)^2
\label{eq:4}
\end{eqnarray}
\end{widetext}
Notice that the expression for total energy only contains the nearest and next-nearest neighbor bond correlation parameters $f_{x}$, $f_{y}$, $g_x$, $g_y$ and $g_{x+y}$. The constraint of zero magnetization, appropriate for $T > T_N$ (for the two dimensional problem $T_N=0$), is enforced by the Lagrange multiplier $\mu$. Minimizing $E-T\mathcal{S} - \mu f(0)$ with respect to
$\varepsilon_{\bf k},\ \phi,\ \theta_{\bf k}$,
we obtain $\tanh 2\theta_{\mathbf{k}}=A_{\mathbf{k}}/B_{\mathbf{k}}$,
$\varepsilon_{\mathbf{k}}=\sqrt{B_{\mathbf{k}}^{2}-A_{\mathbf{k}}^{2}}$
and $\sin \phi \left(f_{y}^{2}+g_{y}^{2}-f_{x}^{2}-g_{x}^{2}\right)=0$,
where
\begin{eqnarray}
&&A_{\mathbf{k}}=2J_1\left(\sin^2 \frac{\phi}{2} g_x \mathcal{C}_{x,\mathbf{k}}+\cos^2 \frac{\phi}{2}g_y \mathcal{C}_{y,\mathbf{k}}\right)\nonumber \\&& \ \ \ \ \ \ \ \ \ \ \ \ \ \ \ \ \ \ \ \ \ \ \ \ \ \ \ \ \ \ \ \ \ \ \ \ \ \ \ \ \ \ \ \ \ \ +4J_2\ g_{x+y}\mathcal{C}_{x+y,\mathbf{k}}\\
&&B_{\mathbf{k}}=2J_1\left(\sin^2 \frac{\phi}{2} (g_x-f_y)+\cos^2 \frac{\phi}{2}(g_y-f_x)\right)\nonumber \\ &&+2J_1\left(\cos^2 \frac{\phi}{2}f_x \mathcal{C}_{x,\mathbf{k}} +\sin^2 \frac{\phi}{2}f_y\mathcal{C}_{y,\mathbf{k}}\right)+4J_2\ g_{x+y}-\mu, \nonumber \\
\end{eqnarray}
and we have introduced the form factors
$\mathcal{C}_{x,\mathbf{k}}=\cos k_{x}a$, $\mathcal{C}_{y,\mathbf{k}}=\cos k_{y}a$, and $\mathcal{C}_{x+y,\mathbf{k}}
=\cos k_xa \cos k_ya$.
Now using $\tanh 2\theta_{\mathbf{k}}=A_{\mathbf{k}}/B_{\mathbf{k}}$ in Eq.~\ref{eq:FAFcorrelators}, we obtain the following set of self-consistent equations

\begin{eqnarray}
&&f_{\alpha}=\frac{1}{N}\sum_{\mathbf{k}}\frac{B_{\mathbf{k}}}
{\epsilon_{\mathbf{k}}}\left(n_{\mathbf{k}}+\frac{1}{2}\right)\mathcal{C}_{{\alpha},\mathbf{k}}, \ {\alpha}=x,y  \label{self-consistent1}\\
&&g_{\alpha}=\frac{1}{N}\sum_{\mathbf{k}}\frac{A_{\mathbf{k}}}{\epsilon_{\mathbf{k}}}
\left(n_{\mathbf{k}}+\frac{1}{2}\right)\mathcal{C}_{{\alpha},\mathbf{k}}, \ {\alpha}=x,y,x+y  \label{self-consistent2}\\
&&S+\frac{1}{2}=f(0)=\frac{1}{N}\sum_{\mathbf{k}}\frac{B_{\mathbf{k}}}
{\epsilon_{\mathbf{k}}}\left(n_{\mathbf{k}}+\frac{1}{2}\right)
\label{self-consistent3}
\end{eqnarray}

We identify two important temperature
scales $T_0$ and $T_{\sigma 0}$ such that $T_0>T_{\sigma 0}$,
by solving the self-consistent equations.
The temperature
$T_0=J_2(S+1/2)[\log(1/S+1)]^{-1}$ marks the onset of the largest bond
correlation
$g_{x+y}$,
while
$T_{\sigma 0}$ marks the onset of nearest-neighbor bond correlations.
For $T>T_0$, all the bond correlations vanish and we have decoupled
local moment
behavior.
The first-order transition  from the correlated to decoupled moment state
at $T_0$ is an artifact of the mean-field
theory \cite{Takahashi1}. In the temperature range $T_{\sigma 0}<T<T_{0}$,
the sublattice
angle $\phi$ remains arbitrary, and the system has $C_{4v}$ rotational
symmetry.
For $T<T_{\sigma 0}$ there are two degenerate solutions
$\phi=\pi$, with
$g_y=f_x=0, g_x\neq0, f_y \neq 0, g_x\neq f_y$,
and $\phi=0$, with $x \leftrightarrow y$ switching.
An Ising order parameter, which is defined classically as
$\sigma=\boldsymbol \Omega_1 \cdot \boldsymbol \Omega_2 = \cos \phi$,
is modified
to $\sigma \propto 2\left(\cos^2\frac{\phi}{2}(f_{x}^{2}+g_{y}^{2})
-\sin^2\frac{\phi}{2}(f_{y}^{2}+g_{x}^{2})\right)$,
and  becomes nonzero
below $T_{\sigma 0}$.
We identify this temperature
as the mean-field ``Ising transition" temperature;
fluctuations will reduce the actual transition to
$T_{\sigma} < T_{\sigma 0}$.
In the following, we will focus
on the state with $\phi=\pi$.
The spectrum is gapped at any nonzero temperature,
but
becomes gapless at $T=0$ giving rise to
$(\pi,0)$ antiferromagnetic order via a Bose condensation.

\section{Low energy spectrum and correlation length}
\begin{figure}[t!]
\centering
\includegraphics[scale=0.315]{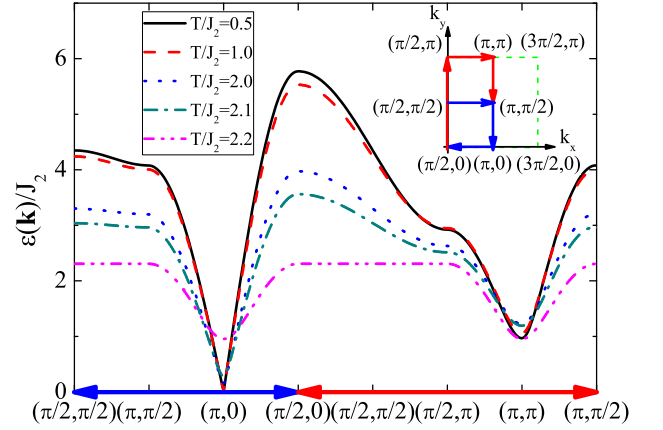}
\caption{(Color online)The dispersion $\varepsilon_{\mathbf{k}}$ along high
symmetry directions in the paramagnetic Brillouin zone for different
temperatures and $S=1$, $J_1/J_2=0.8$.
The curves from top to bottom viewed at the left end are for $T/J_2=0.5, 1.0, 2.0, 2.1,2.2$.
The plotted directions in the
Brillouin zone are displayed in the upper right corner}
\label{fig:2}
\end{figure}

The boson dispersion $\epsilon_{\mathbf{k}}$ is shown in Fig.~\ref{fig:2}.
For $\phi=\pi$, and $T \ll T_{\sigma 0}$, the low energy physics is governed by the excitations in the vicinity of the ordering vector $(\pi,0)$, where the absolute minimum of the dispersion is located. Near $(\pi,0)$,
the dispersion can be approximated by
\begin{eqnarray}
\epsilon_{\mathbf{k}}&=&\left[v_{1x}^{2}(\pi-k_{x})^2+v_{1y}^{2}k_{y}^{2}+\Delta_{1}^{2}\right]^{\frac{1}{2}}\\
\Delta_{1}&=&\left[-\mu (8J_2g_{x+y}+4J_1g_x-\mu)\right]^{\frac{1}{2}}\\
v_{1x}&=&a(4J_2 g_{x+y}+2J_1 g_x) \label{eq:v1x}\\
v_{1y}&=&a\bigg[(4J_2 g_{x+y}+2J_1 g_x)(4J_2g_{x+y}-2J_1f_y)\nonumber \\&& \ \ \ \ \ \ \ \ \ \ \ \ \ \ \ \ \ \ \ \ \ \ \ \ \ \ \ \ \ \ \ \ \ \ \ \ \ \ \ \ \ +2J_1f_y\mu\bigg]^{\frac{1}{2}}\label{eq:v1y}
\end{eqnarray}
Similarly in the vicinity of $(0,\pi)$, the excitation can be approximated as

\begin{eqnarray}
\epsilon_{\mathbf{k}}&=&\left[v_{2x}^{2}k_{x}^2+v_{2y}^{2}(\pi-k_{y})^{2}+\Delta_{2}^{2}\right]^{\frac{1}{2}}\\
\Delta_{2}&=&\left[(8J_2g_{x+y}-4J_1f_y-\mu)(4J_1g_x-4f_y-\mu)\right]^{\frac{1}{2}}\\
v_{2x}&=& a(4J_2 g_{x+y}-2J_1 g_x)\\
v_{2y}&=&a\bigg[4J_2g_{x+y}(4J_2 g_{x+y}-2J_1 g_x)+2J_{1}f_y(4J_2g_{x+y}\nonumber \\ && \ \ \ \ \ \ \ \ \ \ \ \ \ \ \ \ \ \ \ \ \ \ \ \ \ \ \ \ \ \ +2J_1g_x-4J_1f_y-\mu)\bigg]^{\frac{1}{2}}.
\end{eqnarray}

At low temperatures $T\ll T_{\sigma 0}$, the Lagrange multiplier $\mu$ is exponentially small, and $\Delta_1 \ll \Delta_2$. Therefore the spin-spin correlation length at low temperatures will be dominated
by the smallest gap
$\Delta_1 =T\exp[-\Delta_J/T]$, where $\Delta_J=2\pi
\rho$ is the Josephson energy,
with
$\rho=m_0v_{1y}$ being the stiffness
and $m_0$ the staggered magnetization at
$T=0$. The velocity anisotropy yields two correlation lengths,
$\xi_x=v_{1x}/\Delta_1$ and $\xi_y=v_{1y}/\Delta_1$.

The low energy excitations around $(\pi,0)$ can also be described in terms of an anisotropic $O(3)$ nonlinear sigma model. Ignoring the $1/S$ corrections and weak temperature dependence of the bond parameters, we can take $g_x=f_y=g_{x+y}=S$, and obtain bare parameters of the sigma model $\chi_{\perp 0}^{-1}=4(2J_2+J_1)a^2$, $\rho_{x0}=(2J_2+J_1)S^2$, and $\rho_{y0}=(2J_2-J_1)S^2$. The spatial anisotropy is captured by two direction dependent spin stiffness constants $\rho_{x0}$ and $\rho_{y0}$, and $\chi_{\perp 0}$ is the bare uniform transverse susceptibility. The spin wave velocities before $1/S$ corrections are given by $v_{1x}=\sqrt{\rho_{x0}/\chi_{\perp 0}}$, and $v_{1y}=\sqrt{\rho_{y 0}/\chi_{\perp 0}}$. The temperature dependence of the gap is determined by the bare Josephson energy scale
\begin{equation}
\Delta_{J0}=2\pi \rho_0=4\pi J_2S^2\sqrt{1-\frac{J_{1}^{2}}{4J_{2}^{2}}},
\end{equation}
where $\rho_0=\sqrt{\rho_{x0}\rho_{y0}}$ is the bare, geometric mean stiffness constant. For parameter values $S=1$, and $J_1/J_2=0.8$, we find $\Delta_{J0}=11.5J_2$. After solving the mean field equations, we obtain the Josephson energy scale
\begin{eqnarray}
&&\Delta_{J}=\frac{\pi m v_{1y}}{a}=\pi m \bigg[(4J_2 g_{x+y}+2J_1 g_x)(4J_2g_{x+y}\nonumber \\&& \ \ \ \ \ \ \ \ \ \ \ \ \ \ \ \ \ \ \ \ \ \ \ \ \ \ \ \ \ \ \ \ \ \ -2J_1f_y)+2J_1f_y\mu \bigg]^{\frac{1}{2}},
\end{eqnarray}
where $m$ is the staggered magnetization at zero temperature and captures the $1/S$ corrections to $\Delta_J$. For $S=1$, $J_1/J_2=0.8$, we have found
the zero temperature parameters $m=0.83$, $g_{x}=0.96$, $f_y=0.91$, $g_{x+y}=1.07$, and $\Delta_J=10.54J_2$. Note that, at $T=0$, our calculation is consistent with that of
Ref.~\cite{Uhrig}. More details regarding the renormalized $\rho$ and $\Delta_J$ obtained from a sigma model calculation will be discussed in Sec.~V. Above $T_{\sigma 0}$, the nearest neighbor bond correlations vanish, and two gaps become equal, $\Delta_1=\Delta_2=\sqrt{-\mu(8J_2g_{x+y}-\mu)}$. As the $C_{4v}$ symmetry is restored above $T_{\sigma0}$, the velocity anisotropy disappears and $v_{1x}=v_{1y}=v_{2x}=v_{2y}=4J_2g_{x+y}a$.
%

\section{Dynamic structure factor}
The dynamic structure factor is calculated in the modified spin wave theory through the average of the longitudinal and transverse spin structure factors. It is expressed as
\begin{eqnarray}
\mathcal{S}(\mathbf{q},\omega)&=&\frac{1}{N}\sum_{k}\sum_{s,\bar{s}=\pm1}
[\cosh(2\theta_{\mathbf{k}+\mathbf{q}}
-2\theta_{\mathbf{k}})-s\bar{s}]\nonumber \\
& &\times \delta(\omega-s\epsilon_{\mathbf{k}+\mathbf{q}}-\bar{s}
\epsilon_{\mathbf{k}})n^{s}_{\mathbf{k}+\mathbf{q}}n^{\bar{s}}_{\mathbf{k}}
\label{S-q-w1}
\end{eqnarray}
where
$n_{\mathbf{k}}^{+}=n_{\mathbf{k}}+1$ and $n_{\mathbf{k}}^{-}=n_{\mathbf{k}}$.

\begin{figure}[t!]
\centering
\includegraphics[scale=0.95]{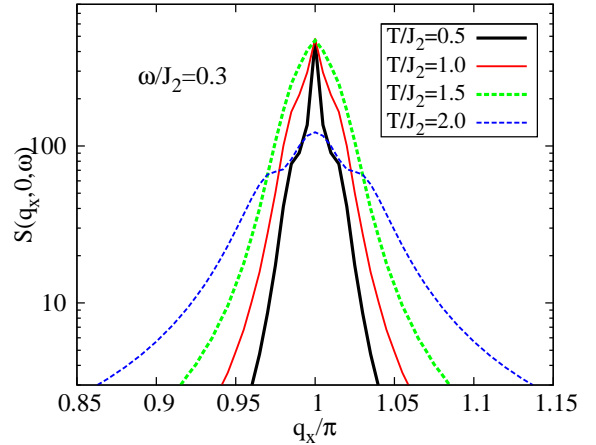}
\caption{(Color online) The sharpening of the dynamic structure
factor around $(\pi,0)$ with decreasing temperature for
$\omega=0.3J_2$, $S=1$, and $J_1/J_2=0.8$.
}
\label{fig:3}
\end{figure}

Consider first $\omega \ll T$, and low temperatures
$T \ll \Delta_J$. The dominant contribution to
$\mathcal{S}(\mathbf{q},\omega)$
comes from the vicinity of the $(\pi,0)$ wave vector.
In the limit $|\pi-q_x|\ll \lambda_{x}^{-1}=T/v_x$
and $q_y \ll \lambda_{y}^{-1}= T/v_y$,
we can analytically \cite{Arovas, Kopietz, Takahashi2}
calculate $\mathcal{S}(\mathbf{q},\omega)$,
which
satisfies a dynamic scaling relation
\begin{eqnarray}
\mathcal{S}(\pi-q_x,q_y,\omega)=\tau
\mathcal{S}_0(\pi-q_x,q_y)\Phi(z,\omega \tau),
\label{S-q-w2}
\end{eqnarray}
where $\mathcal{S}_0(\pi-q_x,q_y)$
is the equal time structure factor, and $\tau=\Delta_{1}^{-1}$ is the scaling time.
${\cal S}_0$ also satisfies a scaling
form $\mathcal{S}_0(\pi-q_x,q_y)=\xi_x\xi_y/(4\pi\lambda_{y}^{2})\Lambda(z)$,
where $z=[(\xi_{x}^{2}(\pi-q_x)^2+\xi_{y}^2q_{y}^2]^{1/2}/2$.
The scaling functions are given by
\begin{eqnarray}
&&\Phi(x,y)=\frac{1}{2\Lambda(x)|y|\sqrt{x^2+(x^2-y^2)^2}}
\bigg(\Theta(x^2-y^2)\frac{2}{\pi}\nonumber \\ && \times \arctan\bigg[|y|
\sqrt{\frac{x^2-y^2}{x^2+(x^2-y^2)^2}}\bigg]+\Theta(y^2-x^2-1)\bigg).\nonumber \\
&&\Lambda(z)=\frac{\log[z+\sqrt{1+z^2}]}{z\sqrt{1+z^2}}
\label{scaling-functions}
\end{eqnarray}
When $z\to 0$, $\Lambda(z) \to 1$,
and for $z \gg 1$, $\Lambda(z) \to \log(z)/z^2$. The second limit corresponds to momentum
scales between inverse correlation length and inverse thermal length,
where the
system appears to have long range order (Goldstone mode behavior). The results
for ${\cal S}_0$ are
in agreement with one loop scaling results of a quantum nonlinear sigma model \cite{Chakravarty, Sachdev}.

\begin{figure}[htbp]
\centering
\label{fig:subfig4b}
\includegraphics[
scale=0.4,
]{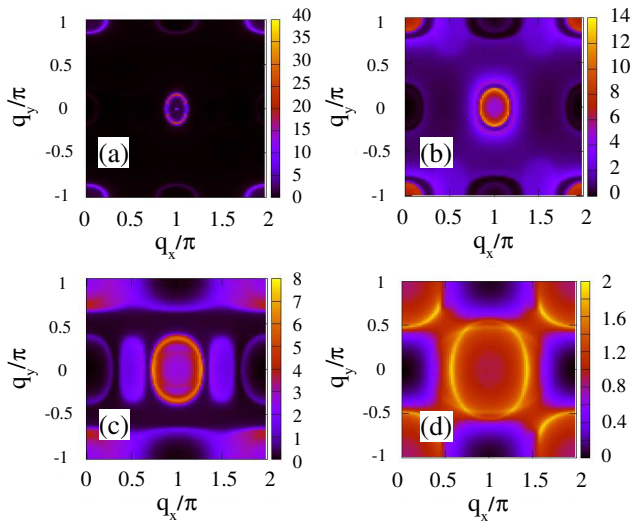}
\label{fig:4}
\caption[]{(Color online)
Distribution of the dynamic structure factor in the
momentum space for different temperatures and energies.
The temperatures and frequencies corresponding to panels (a)-(d) are respectively given by (a) $T/J_2=0.5$, $\omega/J_2=2.0$, (b) $T/J_2=2.1$, $\omega/J_2=2.0$, (c) $T/J_2=2.1$, $\omega/J_2=3.0$, (d) $T/J_2=2.1$, $\omega/J_2=4.5$.}
\end{figure}

A number of features follow from Eqs.~(\ref{S-q-w2},\ref{scaling-functions}).
As a function of energy for a fixed ${\bf q}$ with
$z\gg 1$,
$\mathcal{S}(\mathbf{q},\omega)$ has a broad peak around
$\omega \sim z/\tau$.
As a function of ${\bf q}$ for a fixed $\omega$,
$\mathcal{S}(\mathbf{q},\omega)$ sharpens as temperature
is reduced reflecting the increase of correlation length;
this is also seen from the results of direct numerical
calculations (Fig.~\ref{fig:3}). In the numerical calculations of
$\mathcal{S}(\mathbf{q},\omega)$
in Eq.~(\ref{S-q-w1}), a Lorentzian broadening
of the delta functions has been employed, and consequently
the gap between $\omega \tau <z$ and $\omega \tau >\sqrt{z^2+1}$
is not observed in Fig.~\ref{fig:3} but is
instead left as shoulders. The processes beyond the modified
spin wave theory are expected to smear the two-peak structure
and also modify the scaling time $\tau$ to the phase coherence
time $\sim (\Delta_J/T)^{1/2}/\Delta_{1}$ \cite{Chakravarty, Sachdev}.

Beyond the $\omega \ll T$ limit, we focus
on the distribution of spectral weight in momentum space.
Figs.~4(a) and 4(b) illustrate
the behavior at low energies.
Provided $T < T_{\sigma 0}$, the anisotropy of the correlation
lengths gives rise to an elliptic
feature centered around $(\pi,0)$.
The overall size of the ellipses is reduced as the temperature
is decreased, reflecting increasing correlation lengths.
On the other hand, the ellipticity has only weak temperature
dependence;
the ratio of two correlation lengths
is almost unaffected by temperature variations
for $T\ll T_{\sigma 0}$, due to the weak temperature
dependence of the velocity ratio $v_{1x}/v_{1y}$.

With increasing energy, the evolution of the spectral
weight distribution is illustrated
in Figs.~4(b)-4(d).
At intermediate energies,
when $\omega$ is comparable to the peak energy
in the dispersion $\epsilon_{\mathbf{k}}$ (see Fig.~2),
there are features near $((1\pm 1/2)\pi,0)$, whose spectral
weight is relatively small at the temperature shown in
Fig.~4(c) but will increase with lowering
temperature.
The most visible spectral feature, however, is associated with
the expanding ellipses surrounding $(\pm \pi,0)$ and $(0,\pm \pi)$,
as is clearly seen in the high-energy spectrum shown
in Fig.~4(d).

\section{The role of itinerant electrons and Ginzburg-Landau considerations}

\subsection{Anti-ferromagnetic fluctuations}
The description of the iron pnictides in terms of bad metals invokes
quasi-localized moments coupled to itinerant electrons
whose spectral weight
depends on the proximity of the system to the Mott
transition \cite{Si_NJP}. For the parent compounds, the low-energy
spin dynamics can be described in terms of a Ginzburg-Landau
functional \cite{Si_NJP}
${\cal S}={\cal S}_2+{\cal S}_4 + \ldots $,
where
\begin{eqnarray}
{\cal S}_2&=&\int d {\bf q}
d \omega [( r + w A_{\bf Q} + c {\bf q}^2 +\omega^2
 + \gamma |\omega|)({\bf m}^2+{\bf m'}^2)\nonumber \\
& &+v (q_x^2-q_y^2)
{\bf m} \cdot {\bf m'}],
\label{S-effective}
\end{eqnarray}
where ${\bf m}$ and ${\bf m '}$ are O(3) vectors respectively
for the magnetizations of the two decoupled sublattices,
$q_x$ and $q_y$ are measured with respect to
$(\pm \pi,0)$ or $(0,\pm \pi)$, $w<1$ is the coherent
fraction of the single-electron spectral weight,
and $\gamma$ is the strength of spin
damping caused by the coupling to the itinerant electrons.
${\cal S}_4$ contains not only terms of
the form ${\bf m}^4$, ${\bf m'}^4$ and ${\bf m}^2{\bf m'}^2$,
but also an order-from-disorder term $({\bf m \cdot m'})^2$
with a negative coefficient \cite{Chandra}.
Eq.\ (\ref{S-effective}) implies that elliptic features will
occur in the dynamical responses even in the regime where
the Ising order is not static
but fluctuating
and short-ranged;
the primary role of the itinerant electrons, beyond shifting $r$ through the positive
$w A_{\bf Q}$ term, is to
provide damping effects to such features.

Well below the mean-field Ising transition temperature, the thermal fluctuations of the Ising order parameter $\sigma=\pm\langle \mathbf{m}\cdot \mathbf{m}^{'}\rangle /|\mathbf{m}||\mathbf{m}^{'}|$ in the effective action of Eq.\ (\ref{S-effective})can be ignored. The choice of $\sigma=\pm$ respectively correspond to short range $(\pi,0)$ or $(0, \pi)$ order. For short range $(\pi, 0)$ order, $\mathbf{m}-\mathbf{m}^{'}$ becomes gapped and we find that order parameter dynamics can be approximately determined in terms of a single $O(3)$ order parameter field $\mathbf{M}=\mathbf{m}+\mathbf{m}^{'}$. The effective action for this field at quadratic order is given by
\begin{eqnarray}
\mathcal{S}_2&\approx& T\int d {\bf q}
\sum_{l}\left[r + w A_{\bf Q}  + q_{x}^{2}v_{x}^{2}+q_{y}^{2}v_{y}^{2} +\omega_{l}^{2}
 + \gamma |\omega_l|\right] \nonumber \\ && \ \ \ \ \ \ \ \ \ \ \ \ \ \ \ \ \ \ \ \ \ \ \ \ \ \ \ \ \ \ \ \ \ \ \ \ \ \ \ \ \ \ \ \ \ \ \ \ \ \ \ \ \ \ \ \ \ \ \ \ \ \ \ \ \ \ \ \ \ \times{\bf M}^2
\end{eqnarray}
where $v_{x/y}^{2}= (c\pm v/4)$,
and $\omega_l = 2\pi T l$ is the Matsubara frequency.
With further assumption of small amplitude fluctuations, we can write $\mathbf{M}=M_0\mathbf{n}$, where $M_0$ is the constant amplitude, and $\mathbf{n}$ is the unit vector field. Thus low energy dynamics is now determined by a damped, anisotropic nonlinear sigma model. We consider the following damped nonlinear sigma model action
\begin{equation}
\mathcal{S}_{eff}=\frac{T}{2vg}\int d^{2}q \sum_{l}\left[v^2q^2+\omega_{l}^{2}+\gamma |\omega_l|\right]|\mathbf{n}(\mathbf{q},\omega_l)|^2
\label{eq:sigmamodel}
\end{equation}
In writing the above equation we have rescaled $\sqrt{v_y/v_x}q_x \to q_x$, and $\sqrt{v_x/v_y}q_{y}\to q_y$, and $v=\sqrt{v_xv_y}$, to write the action in spatially isotropic form, and $g=v/\rho=v^{-1}\chi_{\perp}^{-1}$ is the coupling constant with dimension of length. The scaling behavior of the correlation length in the quantum disorder phase and quantum critical regime for this damped nonlinear sigma model has been analyzed in Ref.~\onlinecite{sachdevchubukov}. Here we will only consider the thermally disordered or renormalized classical regime. In the large $N$ limit, the gap in the excitation spectrum $\Delta$ can be determined from the saddle point equation
\begin{equation}
T\sum_{l}\int_{\bar{\Lambda}} \frac{d^2q}{(2\pi)^2}\frac{1}{v^2q^2+\omega_{l}^{2}+\gamma |\omega_l|+\Delta^2}=\frac{1}{vg}
\end{equation}
where $\bar{\Lambda} \sim \pi/a$ is the momentum cutoff. The Matsubara sum can be performed in terms of digamma functions, and after the momentum integration the left hand side can be expressed in terms of the logarithm of the gamma function. Here we consider two extreme limits of $\gamma/(2\pi T) \ll 1$ and $\gamma/(2\pi T) \gg 1$.

In the limit $\gamma/(2\pi T) \ll 1$, we obtain $z=1$ nonlinear sigma model result
\begin{equation}
\sinh \frac{\Delta}{2T}=\sinh \frac{v\bar{\Lambda}}{2T}\exp \left(\frac{2\pi v}{gT}\right)
\end{equation}
In the limit of small temperatures, such that $v\bar{\Lambda} \gg T$, and $\Delta \ll T$, we obtain the result for small $\gamma$ limit,
\begin{eqnarray}
\Delta=T\exp \left(-\frac{2\pi v}{T}\left(\frac{1}{g}-\frac{1}{g_{c1}}\right)\right)=T \exp \left(-\frac{2\pi \rho}{T}\right)
\end{eqnarray}
where $g_{c1}=4\pi/\bar{\Lambda}$ is the coupling strength for zero temperature $z=1$ quantum critical point, and $\rho$ is the renormalized spin stiffness constant. From this expression we find $\xi = v/T\exp(\frac{2\pi \rho}{T})$ in the renormalized classical regime described by $T \ll 2\pi \rho$. For $2\pi \rho \gg T$, one obtains $z=1$ quantum critical behavior $\xi \sim v/T$. If we go beyond the $N\to \infty$ limit, or perform a two loop renormalization group calculation in the renormalized classical regime, we will find the correct classical result $\xi \sim \exp(\frac{2\pi \rho}{T})$\cite{Takahashi2,Chakravarty}
.

For $\gamma/(2\pi T) \gg 1$, the physical properties are governed by a $z=2$ nonlinear sigma model. The frequency sum is performed after imposing a frequency cut-off $\omega_c=v^2\bar{\Lambda}^2/\gamma$, and after performing the momentum integration we obtain
\begin{eqnarray}
&&\frac{2\pi v}{gT}=\log \frac{v\bar{\Lambda}}{\Delta}+\log \Gamma\left(1+\frac{2v^2\bar{\Lambda}^2}{2\pi \gamma T}+\frac{\Delta^2}{2\pi \gamma T}\right)\nonumber \\&&-2\log \Gamma\left(1+\frac{v^2\bar{\Lambda}^2}{2\pi \gamma T}+\frac{\Delta^2}{2\pi \gamma T}\right)+\log \Gamma\left(1+\frac{\Delta^2}{2\pi \gamma T}\right)\nonumber \\
\end{eqnarray}
Now in the limit $2v^2\bar{\Lambda}^2/(2\pi \gamma T) \gg 1$ and $\Delta^2/(2\pi \gamma T) \ll 1$,
 we can use the asymptotic behavior of the $\log \Gamma (1+x)$ to obtain the gap $\Delta$ at large Landau damping,
\begin{equation}
\Delta=v\bar{\Lambda}\exp \left(-\frac{2\pi v}{T}(1/g-1/g_{c2})\right)=v\bar{\Lambda} \exp \left(-\frac{2\pi \rho}{T}\right)
\end{equation}
with $1/g_{c2}=(v\bar{\Lambda}^2)(2\log 2 -1)/(4\pi^2 \gamma) $. Notice that correct renormalized classical behavior of the correlation length for large Landau damping is found from the saddle point equation. For $2\pi \rho \ll T$ we find $z=2$ quantum critical behavior $\Delta \sim \sqrt{2\pi \gamma T}$, augmented by logarithmic corrections. From the expressions of $g_{c1}$, $g_{c2}$ we find that the stiffness for the $z=2$ case is smaller than the $z=1$ case. This reflects the role of Landau damping.

To summarize, in the limits of both small and large Landau damping, the correlation length has an exponential temperature dependence in the renormalized classical regime. 
This will be the basis of our fitting the correlation length, which is described in Appendix A.

If we consider the effects of the inter-layer antiferromagnetic exchange coupling $J_z$ in addition to the $J_1-J_2$ model by using modified spin wave theory (see Appendix C), we obtain a finite mean field anti-ferromagnetic transition temperature $T_{N0}$. Within the Ginzburg-Landau framework this corresponds to setting $r(T)=0$. The fermion contribution $wA_{\mathbf{Q}}$ being positive, will decrease the transition temperature from the mean field value $T_{N0}$ to a smaller value $T_N$. However there will be significant amount of three dimensional antiferromagnetic fluctuations up to the mean field Neel temperature $T_{N0}$. Above $T_{N0}$ the magnetic fluctuations are essentially two dimensional.


\subsection{Ising fluctuations} Since the Ising order parameter breaks $C_{4v}$ symmetry, and in particular corresponds to $B_{1g}$ representation of the tetragonal lattice, it will couple to all the singlet fermion bilinears, which correspond  to $B_{1g}$ representation. Without the loss of generality if we consider a two orbital model of fermions including only $d_{xz}$ and $d_{yz}$ orbitals, the Ising order parameter $\sigma$ will couple to $(\cos k_x -\cos k_y)\Psi_{\mathbf{k}s}^{\dagger}\Psi_{\mathbf{k}s}$, $\Psi_{\mathbf{k}s}^{\dagger}\tau_3\Psi_{\mathbf{k}s}$, $(\cos k_x +\cos k_y)\Psi_{\mathbf{k}s}^{\dagger}\tau_3\Psi_{\mathbf{k}s}$, and $\cos k_x\cos k_y \Psi_{\mathbf{k}s}^{\dagger}\tau_3\Psi_{\mathbf{k}s}$ etc., where $\Psi^{\dagger}_{\mathbf{k}s}=(c^{\dagger}_{xz,\mathbf{k}s}, c^{\dagger}_{yz,\mathbf{k}s})$ describes the orbital and spin dependent fermion creation operators, and the Pauli matrix $\tau_3$ acts on the orbital basis. Among the various $B_{1g}$ bilinears, the conventional nematic order parameter and the ferro-orbital order parameter respectively correspond to $(\cos k_x -\cos k_y)\Psi_{\mathbf{k}s}^{\dagger}\Psi_{\mathbf{k}s}$ and $\Psi_{\mathbf{k}s}^{\dagger}\tau_3\Psi_{\mathbf{k}s}$. Notice that we can couple other $d$ orbitals, following the same symmetry based criterion. When we integrate out the itinerant fermions, the contributions to the Ising order parameter $\sigma$ will arise from generalized $B_{1g}$ particle-hole susceptibilities, and the quadratic part of the low energy action for $\sigma$ will have the form
\begin{eqnarray}
S_{2}[\sigma]=\int d\mathbf{q}\sum_{l}\left[r_{\sigma}+wA_{0}+q^2+\gamma_{\sigma}\frac{|\omega_l |}{q}\right]|\sigma(\mathbf{q},\omega_l)|^2 \nonumber \\
\end{eqnarray}
In the above equation $\gamma_{\sigma}$ is the Landau damping strength, and $r_{\sigma}$ is the mass term arising from the localized model, and $w A_{0}> 0$ is fermion contribution to the Ising mass. This fermionic contribution will suppress the Ising transition temperature from its mean field value $T_{\sigma0}$ to $T_{\sigma}$. But, the correlation length of the Ising order parameter will remain appreciable up to the mean field temperature $T_{\sigma 0}$. Since Ising transition occurs due to in plane magnetic fluctuations, consideration of inter-layer coupling does not significantly modify the Ising correlations.

When we consider the magnetic and Ising order parameter fluctuations on the same footing, further changes in the transition temperatures will arise from the self interaction of $\sigma$, $\mathbf{m}$, $\mathbf{m}^{'}$, and their mutual interaction $\sigma \mathbf{m} \cdot \mathbf{m}^{'}$.
The interplay of Ising and magnetic order parameters, and their self-interactions are crucial to determining if there will be a concomitant first order transition or two separate second order phase transitions. Despite the suppression of actual transition temperatures and the possible complexity regarding the actual nature of the transitions, we still expect that the correlation lengths of the magnetic and the Ising order parameters will remain sizable up to their respective mean-field transition temperatures.

\section{Implications for iron pnictides}
\begin{figure}[ht]
\centering
\includegraphics[
scale=0.6
]
{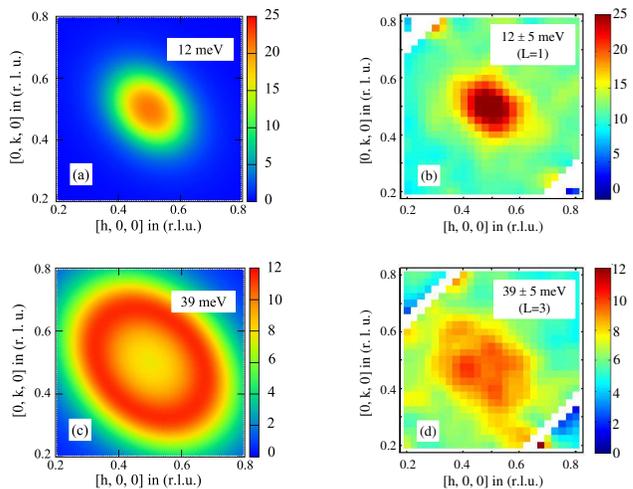}
\label{fig:5}
\caption[]{
Panels (a) and (b) respectively demonstrate $\mathcal{S}(\mathbf{q}-\mathbf{Q},\omega=12 meV)$ at $T=180 K$ obtained from our theory and data of Ref.~\onlinecite{Diallo}. Panels (c) and (d) respectively demonstrate $\mathcal{S}(\mathbf{q}-\mathbf{Q},\omega=39 meV)$ at $T=180 K$ obtained from our theory and data of Ref.~\onlinecite{Diallo}.
We have used
$J_1/J_2=0.55$  \cite{Diallo}, $J_2=9.8meV$
and $\gamma=47 meV$. To facilitate the comparison with experimental result, we have plotted here in
the
 Brillouin zone corresponding to the two-Fe unit cell instead of that for the one-Fe unit cell used
 in the rest of the paper.
}
\end{figure}
Our detailed theoretical studies provide the basis to understand the
anisotropic spin responses that have been observed in the paramagnetic
phase of the parent iron pnictides $\mathrm{CaFe}_2\mathrm{As}_2$ \cite{Diallo}.
These observations, made at temperatures above the first order
antiferromagnetic/structural transition, can be understood
if the transition temperature is
assumed to be considerably lower than the mean-field Ising
transition temperature by the effects of fluctuations and
coupling to phonons. To compare our theoretical results with the experiments of Ref.~\onlinecite{Diallo} we have fitted the low frequency experimental data with the dynamic structure factor calculated within the saddle point approximation of an anisotropic, damped nonlinear sigma model, which follows from the action of Eq.\ (\ref{S-effective}). Within the saddle point approximation the imaginary part of the staggered susceptibility is given by
\begin{equation}
\chi^{''}(\mathbf{q}-\mathbf{Q},\omega)=\frac{\chi_{\perp}^{-1}\gamma \omega}{\gamma^2\omega^2+(\omega^2-v_{x}^{2}(q_{x}-\pi)^{2}-v_{y}^{2}q_{y}^{2}-\Delta^2)^2}
\label{imaginarychi}
\end{equation}
The velocities of the effective model are taken from the modified spin wave calculations. The details of
our procedure are provided in Appendix A.

The comparison of our results with that of Ref.~\onlinecite{Diallo} are shown in Fig.~5. The calculated elliptic features of $\mathcal{S}(\mathbf{q}-\mathbf{Q},\omega)$ (Fig.~5(a)) is compatible with that seen experimentally (Fig.~5(b)) at low frequencies. This continues to be the case at higher frequencies,
as
shown in Fig.~5(c) and Fig.~5(d). The experimental results in the paramagnetic phase are consistent with our conclusions that
as temperature is lowered,
the peaks
in the momentum space sharpen
but the ellipticity
is only weakly affected.
Our estimated values of exchange constants
 are consistent with that of Ref.~\onlinecite{Diallo}. When $\omega$ is smaller than the excitation gap $\Delta$, the dynamic structure factor is peaked at $\mathbf{q}=\mathbf{Q}$. For $\omega>\Delta$, the intensity peak gets shifted to $|\mathbf{q}-\mathbf{Q}|=\sqrt{\omega^2-\Delta^2}/v$ as shown in Fig.~5(c), and the $\omega^2$ term in the dynamics is important to capture this feature also observed in the experiment as shown in Fig.~5(d).

Inter-layer magnetic couplings in the parent iron arsenides vary considerably among the materials, but are always relatively weak. In Ref.~\onlinecite{Diallo}, the inter-layer coupling $J_z$ in paramagnetic CaFe$_{\rm 2}$As$_{\rm 2}$ was shown to be very weak, with $J_z/J_2=0.1$, being smaller than its counterpart in the magnetically ordered phase at low temperature. Consideration of such a weak inter-layer coupling does not appreciably change the estimated exchange constants and the in-plane spin dynamics. An estimation of the spin stiffness constant using a renormalized classical approximation for the correlation length shows that both fermion induced moment reduction, and Landau damping can sufficiently renormalize the stiffness constant (see Appendix A). In Appendix B we have considered the effects of the weak inter-layer exchange coupling $J_z$ using the modified spin wave theory. For $J_z/J_2=0.1$ the mean field Neel temperature $T_{N0}$ and the mean field Ising transition temperature $T_{\sigma 0}$ become very close. However as we have discussed in Sec. V, despite the suppression of the actual transition temperature due to various fluctuation mechanisms, the magnetic and the Ising correlation lengths remain sizable up to the mean-field transition temperatures. In the temperature regime $T_N<T<T_{N0}$, there are three-dimensional antiferromagnetic fluctuations. However if we consider the ratio of the in-plane and inter-plane correlation lengths (measured in units of corresponding lattice spacing), we find $\xi_z/\xi_x \approx (J_z/(2J_2+J_1))^{1/2}$. This ratio is of course material dependent. For weak inter-layer coupling of Ref.~\onlinecite{Diallo}, this ratio is $\sim 0.2$, and magnetic fluctuations are indeed quasi-two dimensional.

Finally our discussion regarding the effect of itinerant electrons
is most pertinent to the parent systems, but is consistent
with the experimental observation
of similar low-energy anisotropic responses in the carrier-doped
iron pnictides \cite{Lester,Li,Park}.

\section{Summary and conclusions} We have addressed the spin dynamics in the paramagnetic phase of a two dimensional $J_1-J_2$ antiferromagnet on a square lattice at a finite temperature, using modified spin wave theory. Within the modified spin wave theory we have identified a mean field Ising transition temperature $T_{\sigma0}$, below which the $C_{4v}$ symmetry of the square lattice is spontaneously broken. In the Ising ordered phase the system demonstrates short range $(\pi,0)$ or $(0,\pi)$ antiferromagnetic order. In order to systematically understand the finite temperature spin-dynamics in the paramagnetic phase of iron pnictides, we have described the fermionic contributions and self-interaction effects of the order parameter fields within a Ginzburg-Landau framework. We have found that the fermion contribution and the self-interaction effects can considerably decrease the Neel and the Ising transition temperatures from their corresponding mean field values. However the correlation lengths of the magnetic and Ising order parameters can remain appreciable up to the mean field transition temperatures. Based on this assumption, we have fitted the experimental data of Ref.~\onlinecite{Diallo}, using our theoretical results. The calculated anisotropic features of the spin response are compatible with experiments for different frequencies.

Finally, our calculations of the spin fluctuations at high energies
should  help understand future experiments. High-energy spin spectrum
at the low-temperature ordered state of CaFe$_{\rm 2}$As$_{\rm 2}$
\cite{Zhao} has already provided valuable information on the
x-y anisotropy of the exchange interactions. Similar experiments
have recently been reported in BaFe$_{\rm 2}$As$_{\rm 2}$ \cite{Harriger}
and SrFe$_{\rm 2}$As$_{\rm 2}$ \cite{Ewings}, including at temperatures
just above the N\'eel transition where strong orbital anisotropy has
developed \cite{Yi,JHChu}. It will be instructive to experimentally map out
the high-energy spectrum at higher temperatures in the paramagnetic phase.

\acknowledgements

We thank S. Chakravarty, P. Dai, R. J. McQueeney, and A. Nevidomskyy
for valuable discussions, and
NSF Grant No. DMR-1006985, the Robert A. Welch Foundation
Grant No. C-1411 and the W.\ M.\ Keck Foundation for support.

\begin{figure}[t!]
\centering
\includegraphics[scale=0.35]{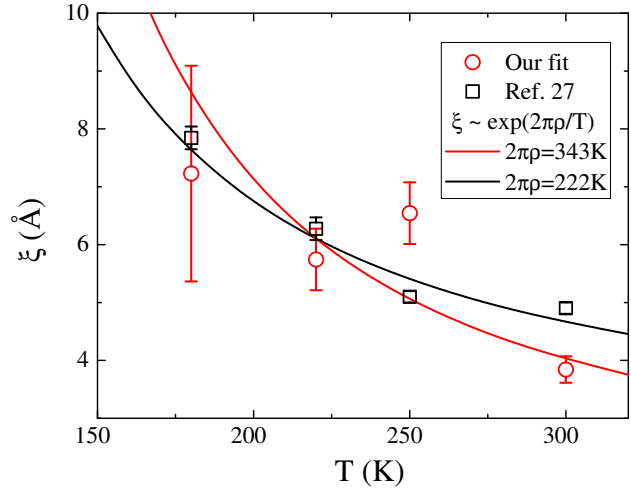}
\caption{(Color online)
Comparison between
the correlation lengths $\xi=\sqrt{\xi_x\xi_y}$ extracted from our fitting and that of Ref.~\onlinecite{Diallo}.
The fitted spin stiffness constants
are shown in the inset.}
\label{fig:6}
\end{figure}

\appendix
\section{ Procedure of comparing with the experimental data} We analyze the experimental data of Ref.~\onlinecite{Diallo} by using the imaginary part of the staggered susceptibility $\chi(\mathbf{q}-\mathbf{Q},\omega)$, which is calculated within the saddle point approximation for Eq.~\ref{eq:sigmamodel}. From the saddle point calculation we find
\begin{equation}
\chi^{''}(\mathbf{q}-\mathbf{Q},\omega)=\frac{\chi_{\perp}^{-1}\gamma \omega}{\gamma^2\omega^2+(\omega^2-v_{x}^{2}(q_{x}-\pi)^{2}-v_{y}^{2}q_{y}^{2}-\Delta^2)^2}
\label{imaginarychi1}
\end{equation}
At $\mathbf{q}=\mathbf{Q}$, we have
\begin{equation}
\chi^{''}(0,\omega)=\frac{\chi_{\perp}^{-1}\gamma \omega}{\gamma^2\omega^2+(\omega^2-\Delta^2)^2}
\label{imaginarychi2}
\end{equation}
where $\chi_{\perp}$ is the uniform transverse susceptibility. We calculate the velocities $v_x$ and $v_y$ using Eq.~\ref{eq:v1x} and Eq.~\ref{eq:v1y}, and for the exchange constants we choose $J_2=10 meV$ and $J_1=0.55J_2=5.5 meV$, as determined by Diallo {\it et al.} \cite{Diallo}. By fitting the experimental data we determine the temperature independent Landau damping strength $\gamma$ and the temperature dependent gap $\Delta$. By fitting the data for $\chi^{''}(0,\omega)$ at $T=180 K$,
with the formula from Eq.~\ref{imaginarychi2}, we find the Landau damping strength $\gamma$ and the gap $\Delta$ at $180K$.
At low frequencies, Eq.~(\ref{imaginarychi2}) can be further approximated by a Lorentzian with a
width
$\Gamma_T \approx \Delta^2/\sqrt{\gamma^2-2\Delta^2}$.
At the relatively low temperature of $180K$, the Lorentzian form is a good fit to Eq.~(\ref{imaginarychi2}) up to frequencies of about $40$ meV, and $\Gamma_T=7meV$.
At the high temperature of $300K$, the Lorentzian form, which becomes a poorer fit to Eq.~(\ref{imaginarychi2}) over the same frequency range, yields $\Gamma_T=44meV$.
The definition of the energy line-width as $\Gamma_T$
is the same notation as used in
Ref.~\onlinecite{Diallo}, but the constant $\gamma$ used in Ref.~\onlinecite{Diallo} is not the conventional Landau damping strength and has a different meaning
from ours. Our estimation
 is $\gamma=47 meV$. For the available data at nonzero $\mathbf{q}-\mathbf{Q}$ at different temperatures, we use the value of
$\gamma$ so determined,
and find the $\Delta$ at different temperatures. Using the values of $v_x$, $v_y$ and $\Delta$, we find the correlation length $\xi_{x}$ and $\xi_y$. The comparison of our theoretically calculated dynamic structure factor with the fitted parameter values, and the experimental results at low frequency $12 meV$ are shown in Fig. 5(a) and Fig. 5(b) of the main text. Even at higher energy $\omega=39 meV$ our results for the dynamic structure factor are in reasonable agreement with experimental data, and the comparison for this frequency is shown in Fig.~5(c) and Fig.~5(d) of the main text. The consideration of the $\omega^2$ term in the effective action, leads to an interesting feature of the dynamic structure factor. For low frequencies such that $\omega<\Delta$, $\chi^{''}(\mathbf{q}-\mathbf{Q},\omega$ is peaked at $\mathbf{q}=\mathbf{Q}$. But at higher frequencies such that $\omega > \Delta$, the intensity peak occurs away from the antiferromagnetic wave-vector and its location is determined by $|\mathbf{q}-\mathbf{Q}|=\sqrt{\omega^2-\Delta^2}/v$. This shift in the intensity peak can clearly seen by comparing Fig.~5(a) and Fig.~5(c). Similar shift in the intensity peak can also be seen in the experimental results by comparing Fig.~5(b) and Fig.~5(d). For the correlation length we have compared our and experimental results in Fig.~6 by plotting the temperature dependence of the geometric mean of $\xi_x$ and $\xi_y$. By fitting the correlation length with the renormalized classical formula, we have obtained an estimation for the stiffness constant. 
The fermion induced reduction of the magnetic moment $M_0$, and Landau damping are found to significantly reduce the stiffness constant in comparison to a pure $J_1-J_2$ model.

\section{Effects of inter-layer exchange coupling} The quasi-2D nature of the spin dynamics was clearly shown in Ref.~\onlinecite{Diallo}. To explain the observed $(\pi, 0, \pi)$ antiferromagnetic order, an inter-layer antiferromagnetic coupling $J_z$ was assumed and $J_z$ was estimated to be $\sim 0.1 J_2$. To assess the effects of $J_z$ on the spin dynamics we first incorporate the three-dimensional effects in our modified spin wave theory calculations. For simplicity we assume the sublattice angle $\phi=\pi$. The modification to our discussion in Sec. I comes through an additional inter-layer antiferromagnetic bond correlation parameter $g_z$. The Ising transition will be determined by the vanishing of in plane nearest-neighbor bond correlations $g_x$ and $f_y$. In the presence of $J_z$, there is a finite, mean-field antiferromagnetic transition temperature $T_{N0}$, corresponding to Bose condensation of $a$'s. The expression for total energy in Eq.~\ref{eq:4} changes into

\begin{eqnarray}
E=-\frac{J_1N}{2}\sum_{\boldsymbol \delta_1=\pm \hat{x}}\bigg(S+\frac{1}{2}-f(0)+g_x\bigg)^2 \nonumber \\
+\frac{J_1N}{2}\sum_{\boldsymbol \delta_2=\pm \hat{y}}\bigg(S+\frac{1}{2}-f(0)+f_y\bigg)^2 \nonumber \\-\frac{J_2N}{2}\sum_{\boldsymbol \delta_3=\pm \hat{x}\pm
\hat{y}}\bigg(S+\frac{1}{2}-f(0) +g_{x+y}\bigg)^2\nonumber \\-\frac{J_zN}{2}\sum_{\boldsymbol \delta_1=\pm \hat{z}}\bigg(S+\frac{1}{2}-f(0)+g_x\bigg)^2,
\end{eqnarray}
and the expressions for $A_{\mathbf{k}}$ and $B_{\mathbf{k}}$ are modified according to

\begin{eqnarray}
A_{\mathbf{k}}&=&2J_1g_x \mathcal{C}_{x,\mathbf{k}}+4J_2\ g_{x+y}\mathcal{C}_{x+y,\mathbf{k}}+2J_z g_z\mathcal{C}_{z,\mathbf{k}}\\
B_{\mathbf{k}}&=&2J_1(g_x-f_y)+2J_1f_y\mathcal{C}_{y,\mathbf{k}}+4J_2\ g_{x+y}-\mu+2J_zg_z,\nonumber \\
\end{eqnarray}
where $\mathcal{C}_{z,\mathbf{k}}=\cos k_z c$, and $c$ is the inter-layer separation. After accounting for the possibility of a finite staggered magnetization below $T_{N0}$, the mean-field equations are given by
\begin{eqnarray}
&&f_{y}=m_0+\frac{1}{N}\sum_{\mathbf{k}}^{'}\frac{B_{\mathbf{k}}}
{\epsilon_{\mathbf{k}}}\left(n_{\mathbf{k}}+\frac{1}{2}\right)\mathcal{C}_{y,\mathbf{k}},\\
&&g_{\alpha}=m_0+\frac{1}{N}\sum_{\mathbf{k}}^{'}\frac{A_{\mathbf{k}}}{\epsilon_{\mathbf{k}}}
\left(n_{\mathbf{k}}+\frac{1}{2}\right)\mathcal{C}_{{\alpha},\mathbf{k}}, \ {\alpha}=x,x+y,z \nonumber \\ \label{self-consistent2}\\
&&S+\frac{1}{2}=m_0+\frac{1}{N}\sum_{\mathbf{k}}^{'}\frac{B_{\mathbf{k}}}
{\epsilon_{\mathbf{k}}}\left(n_{\mathbf{k}}+\frac{1}{2}\right)
\label{self-consistent3}
\end{eqnarray}

\begin{figure}[t!]
\centering
\includegraphics[scale=0.95]{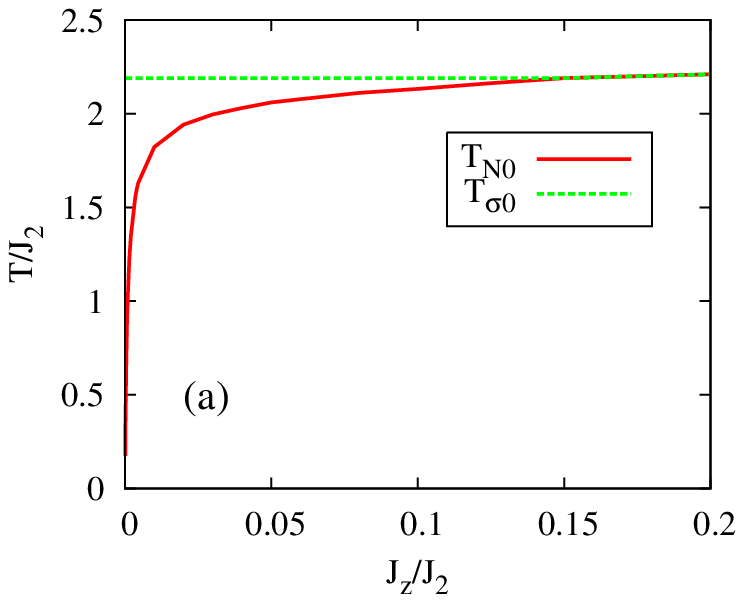}
\label{fig:7a}
\includegraphics[scale=0.94]{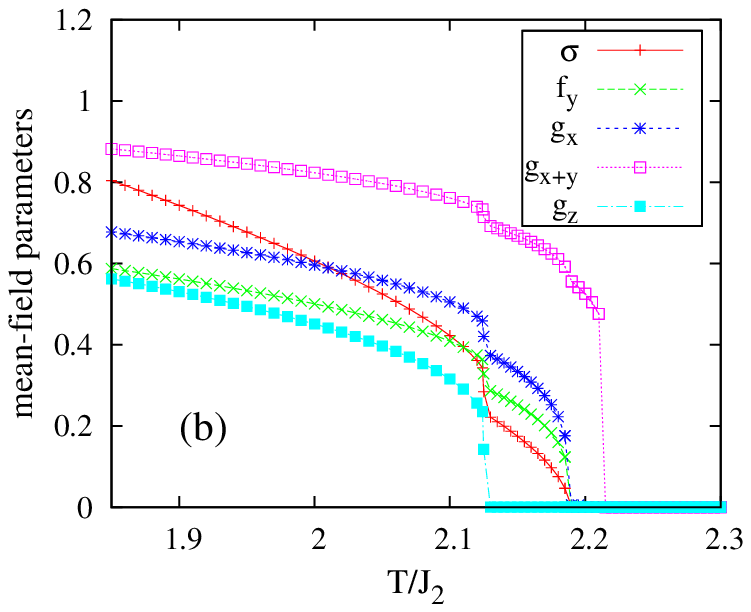}
\label{fig:7b}
\label{fig:7}
\caption[]{Panel (a) shows the comparison between the mean-field Neel temperature $T_{N0}$ and mean-field Ising transition temperature $T_{\sigma 0}$, as a function of the inter-planar coupling $J_z$, for $J_1/J_2=0.8$, and $S=1$. Panel (b) shows the temperature dependence of different mean-field parameters for $J_z/J_2=0.1$.
}
\end{figure}

For $S=1$, $J_1/J_2=0.8$, $c=a$, the dependence of $T_{N0}$ and $T_{\sigma 0}$ on $J_z/J_2$ are shown in Fig.~7(a). The temperature dependence of the mean-field bond parameters for $J_z/J_2=0.1$ are shown in Fig.~7(b). With increasing $J_z$, the Neel temperature gradually increases and asymptotically approaches $T_{\sigma0}$. Since the mean-field Ising transition is a consequence of the two-dimensional magnetic fluctuations, $T_{\sigma 0}$ is not modified by the finite inter-layer coupling $J_z$. For $J_z/J_2=0.1$, and and $J_2\sim 10 meV$ we obtain $T_{N0} \approx T_{\sigma 0} \sim 240 K$, which is much higher than the actual Neel and structural transition temperature. Therefore the fluctuating anisotropy effects will be important over a wide range of temperature, and the finite $J_z$ does not change this conclusion.

Below $T_{\sigma 0}$, by expanding the dispersion around $\mathbf{Q}=(\pi, 0, \pi)$, we obtain

\begin{eqnarray}
\epsilon_{\mathbf{k}}&=&\left[v_{x}^{2}(\pi-k_{x})^2+v_{y}^{2}k_{y}^{2}+v_{z}^{2}(\pi-k_z)^2+\Delta^{2}\right]^{\frac{1}{2}}\\
\Delta&=&\left[-\mu (8J_2g_{x+y}+4J_1g_x+4J_zg_z-\mu)\right]^{\frac{1}{2}}, \nonumber \\ &&  \ \ \ \ \ \ \ \ \ \ \ \ \ \ \ \ \ \ \ \ \ \ \ \ \ \ \ \ \ \ \ \ \ \ \ \mu=0, \mathrm{for} \ \ T<T_N \\
v_{x}&=&a\bigg[(4J_2 g_{x+y}+2J_1 g_x)(4J_2 g_{x+y}+2J_1 g_x \nonumber \\ && \ \ \ \ \ \ \ \ \ \ \ \ \ \ \ \ \ \ \ \ \ \ \ \ \ \ \ \ \ \ \ \ \ \ \ \ \ \ \ \ \ \ \ \ \ \ \ \ \ \ \ \ \ \ +2J_zg_z)\bigg]^{\frac{1}{2}} \\
v_{y}&=& a\bigg[(4J_2 g_{x+y}+2J_1 g_x+2J_zg_z)(4J_2g_{x+y}-2J_1f_y)\nonumber \\ && \ \ \ \ \ \ \ \ \ \ \ \ \ \ \ \ \ \ \ \ \ \ \ \ \ \ \ \ \ \ \ \ \ \ \ \ \ \ \ \ \ \ \ \ \ \ \ \ \ \ \ \ \ \ \ \ +2J_1f_y\mu\bigg]^{\frac{1}{2}} \\
v_{z}&=& c\left[(4J_2 g_{x+y}+2J_1 g_x+2J_zg_z)2J_zg_z\right]^{\frac{1}{2}}
\end{eqnarray}

We further notice that the velocities are well approximated by
\begin{eqnarray}
&&v_x \approx 2Sa(J_1+2J_2)\sqrt{1+\frac{J_z}{J_1+2J_2}}\\
&&v_y \approx v_x \sqrt{\frac{2J_2-J_1}{2J_2+J_1}}\\
&&v_z \approx v_x \frac{c}{a} \sqrt{\frac{J_z}{2J_2+J_1}}
\end{eqnarray}
and even in the presence of finite $J_z$, the ratio $v_y/v_x$ remains unchanged. For $J_z/J_2=0.1$ and $c/a \approx 3.026$ \cite{Diallo} we obtain $v_z/v_x \sim 0.6$, and this leads to smaller inter-planar correlation length ($\xi_z<\xi_{x}, \sqrt{\xi_{x}\xi_{y}}$). In our comparison with experiments we have looked at the data that corresponds to in-plane dynamics, i.e., $\mathbf{q}-\mathbf{Q}=(q_x-\pi,q_y,0)$, and consequently all the formulas remain unaffected. We also note that the effects of inter-planar coupling inside the magnetically ordered phase have been considered in Refs.~\cite{Smerald, Holt, Ong} using similar technique. However our results are derived for the paramagnetic phase, which are essentially different from those described in Refs.~\cite{Smerald, Holt, Ong}.

\end{document}